\documentclass[aps,amsfonts,nofootinbib,a4paper,twocolumn]{revtex4}
\usepackage{graphicx}
\usepackage{amssymb}

\newcommand {\eq}{\begin{equation}}
\newcommand {\ee}{\end{equation}}

\begin{document}

\title{Reconstructing f(R) theory according to holographic dark
energy}

\author{Xing Wu and Zong-Hong Zhu\footnote{zhuzh@bnu.edu.cn}}

\address {Department of Astronomy, Beijing Normal University,
Beijing 100875, People's Republic of China}

\begin{abstract}
In this paper a connection between the holographic dark energy model
and the $f(R)$ theory is established. We treat the $f(R)$ theory as
an effective description for the holographic dark energy and
reconstruct the function $f(R)$ with the parameter $c>1$, $c=1$ and
$c<1$, respectively. We show the distinctive behavior of each cases
realized in $f(R)$ theory, especially for the future evolution.
\end{abstract}
\maketitle

\section{Introduction}
Since 1998, the Type Ia Supernovae observations\cite{SN} have
indicated that the expansion of the universe is currently
accelerating. This result has then been further confirmed by
independent observations of Cosmic Microwave Background
(CMB)\cite{CMB} and Large Scale Structure (LSS)\cite{LSS}. One
explanation for the cosmic acceleration is ascribed to adding an
exotic energy component with negative pressure, dubbed the dark
energy, of which the origin and nature is still a mystery. Various
dark energy models have been proposed in the literature
(see\cite{0603057} for a detailed review). Among others, the
simplest candidate for dark energy is the cosmological constant or
the vacuum energy. Fitted quit well with observational data though
it is, this model suffers from the famous cosmological constant
problem\cite{cc}. This problem arises primarily due to the fact that
the vacuum energy is considered within the framework of quantum
field theory in Minkowski background. As we known, however, at
cosmological scales where the effect of gravity has to be taken into
account, the above description of the vacuum energy would break
down, and it is believed that the correct theoretical value of the
vacuum energy will be predicted by a complete theory of quantum
gravity. Although we are far from reaching such a fundamental
theory, we do know some features of it. The holographic
principle\cite{holo} is an importance feature which can shed some
light on the cosmological constant problem and the dark energy
problem. According to this principle, considering gravity, the
number of the degree of freedom of a local quantum field theory
system is related to the area of its boundary, rather than the
volume of the system as expected when gravity is absent. Along this
line, Cohen et al.\cite{Cohen} suggested an entanglement relation
between the IR and UV cut-offs due to the limitation set by the
formation of a black hole, which in effect sets an upper bound for
the vacuum energy \eq L^3\rho_\Lambda<LM_p^2, \ee where
$\rho_\Lambda$ is related to the UV cut-off, $L$ is the IR cut-off
and $M_p$ is the reduced Planck mass. The form of dark energy is
proposed by saturating the bound as \eq \rho_\Lambda={3c^2M_p^2\over
L^2},\label{holo vac}\ee where $c$ is a numerical factor. It is easy
to check that insert $L=H_0^{-1}$ may give rise to a energy density
compatible with current observation in orders of magnitude. However,
as Hsu\cite{Hsu} pointed out, this can not lead to a desired
equation of state. Li\cite{Li} proposed the holographic dark energy
model, where $L$ is chosen to be the future event horizon \eq
R_{eh}=a\int_t^\infty
\frac{dt'}{a(t')}=a\int_a^\infty\frac{da'}{Ha'^2}. \label{eh}\ee
This model has been tested to be well consistent with current
observations\cite{holo fit}, and it is a compelling candidate for
solving the dark energy problem.

Adding new component of dark energy to the whole energy budget is
one way, there are also other promising ways without resorting to
new forms of energy. Since general relativity is only tested within
solar system up to now, we may well consider the modification to the
Einstein-Hilbert action in Einstein gravity at larger scales with
higher order curvature invariant terms such as $R^2,\
R^{\mu\nu}R_{\mu\nu},\ R^{\mu\nu\alpha\beta}R_{\mu\nu\alpha\beta},\
{\rm or}\ R\Box^kR$ as well as nonminimally coupled scalar fields
with terms like $\phi^2R$. Furthermore, these terms naturally emerge
as quantum corrections in the low energy effective action of quantum
gravity or string theory\cite{effective action QG, effective action
string}. Here we focus on the f(R) theories where the modification
is a function of the Ricci scalar only. In\cite{Starobinsky}, the
author first introduced an additional $R^2$ to the Einstein-Hilbert
action leading to an inflationary solution for the early universe.
As for applications to the dark energy problem, it is shown that
adding an term $1/R^n$\cite{Carroll} or more generalized
$c_1/R^n+c_2R^m$ \cite{R-nRm} can lead to late time acceleration
originated from pure geometrical effect, equivalent to introducing
an effective dark energy in the Einstein frame (see, for example,
Sec. XVI. in \cite{0603057} and the references therein for more
works dedicated to solving the dark energy problem with $f(R)$
theories).

It can be shown\cite{conformal} that through conformal
transformations, extended theories with higher order terms and/or
nonminimally coupled scalar fields correspond to Einstein gravity
with some minimally coupled scalar fields(quintessence) suggesting,
to some extent, an equivalence between dynamic dark energy models
and f(R) theories. An approach was proposed in\cite{fR recon} to
reconstruct the form of $f(R)$ from a given expansion history
$H(z)$. From observational data such as SNe distance modulus vs
redshift, we can obtain the luminosity distance $D_L(z)$ through
which $H(z)$ is given by\eq
H(z)=\left\{\frac{d}{dz}\left[\frac{D_L(z)}{1+z}\right]\right\}^{-1}.\ee
In fact, however, large errors in current observational data prevent
using this procedure to determine the exact form of $f(R)$. But the
important thing is that we can use a given $H(z)$ predicted from a
dark energy model to reconstruct its equivalent $f(R)$ theory. For
example, the $f(R)$ theories reconstructed from the $H(z)$ given by
quiessence and the Chaplygin gas respectively were obtained
in\cite{fR recon}. In this paper, we treat the holographic dark
energy model as one inspired by the holographic principle, an
important feature of a more fundamental theory of quantum gravity,
and to reconstruct the corresponding $f(R)$ theory as an equivalent
description. Compared with previous works, where the holographic
dark energy is reconstructed within scalar field models like ghost
condensate\cite{holo quintom}, quintessence\cite{holo quintessence},
tachyon\cite{holo tachyon} and hessence\cite{holo hessence}, here we
perform the reconstruction in $f(R)$ theory without resorting to any
additional dark energy component, that is, the holographic dark
energy is effectively described by the modification of gravity.

In addition we note that there are in fact two strategies in f(R)
theories: the metric formalism, where the action is varied with
respect to the metric only; and the Palatini
formalism\cite{Palatini}, where the metric and the connnection are
treated as two independent variables with respect to which the
action is varied. It is only in Einstein gravity $f(R)=R$ that both
approaches reach the same result. In general f(R) theories, the
problem of which approach should be used is still an open question
and the final solution may be determined by further observations and
theoretical development. At present we assume the metric formalism
in this paper.

\section{Reconstruction of $f(R)$ theory}
Now let's consider a homogeneous and isotropic universe with flat
spatial geometry consisting of matter and the vacuum energy given by
(\ref{holo vac}). The Friedmann equation reads \eq
3M_p^2H^2=\rho_m+\rho_\Lambda,\ee where $\rho_m=\rho_{m0}(1+z)^3$ by
the equation of energy conservation, and a subscript $0$ denotes the
value at present. By introducing $\Omega_\Lambda={\rho_\Lambda\over
3M_p^2H^2}$ and $\Omega_m={\rho_m\over
3M_p^2H_0^2}=\Omega_{m0}(1+z)^3$, we obtain the Hubble parameter \eq
H(z)=H_0\sqrt{\Omega_m\over 1-\Omega_\Lambda}\label{H(z)}.\ee
Clearly, once we determine the evolution of $\Omega_\Lambda(z)$, the
whole expansion history $H(z)$ is determined. Combining the
definitions of holographic dark energy and the event horizon
(\ref{holo vac}) and (\ref{eh}) we get \eq
\int_a^\infty\frac{da'}{Ha^2}={c\over Ha\sqrt{\Omega_\Lambda}}.\ee
with the initial condition given by setting $z=1$ in (\ref{H(z)})
\eq \Omega_{m0}+\Omega_{\Lambda 0}=1. \label{initial}\ee Inserting
(\ref{H(z)}) into the above equation and taking derivative with
respect to $z$ on both sides(using $1+z=1/a$), we obtain the
differential equation of $\Omega_\Lambda$ \eq
\Omega'_\Lambda=-{1\over
(1+z)}\Omega_\Lambda(1-\Omega_\Lambda)(1+{2\over
c}\sqrt{\Omega_\Lambda}),\label{domega}\ee where the prime denotes
derivative with respect to $z$. We can see that $c$ is the only
parameter affecting the dynamics of the holographic dark energy. In
fact, we can use the equation of energy conservation of the dark
energy to get the equation of state of the dark energy \eq
w_\Lambda=-{1\over3}-{2\over c}\sqrt{\Omega_\Lambda}.\ee It is easy
to see that when $\Omega_\Lambda\rightarrow 1$ in the future, for
$c>1$ the EoS will always be greater than $-1$ behaving like a
quintessence; for $c=1$ the universe will end up with a de Sitter
phase; and for $c<1$ the universe will end up with a phantom phase
and the EoS crossing $-1$ occurs during the evolution exhibiting a
quintom-like behavior. Therefore the parameter $c$ plays a very
important role in determining the evolutionary nature of the
holographic dark energy. Many works\cite{holo fit} have been devoted
to constrain this parameter by observations such as SNe, CMB and
galaxy clusters etc. Almost all the best fits indicate $c<1$,
although $c>1$ is also compatible with the data within $1\sigma$.

\begin{figure}[htbp]
\begin{center}
\includegraphics[scale=0.45]{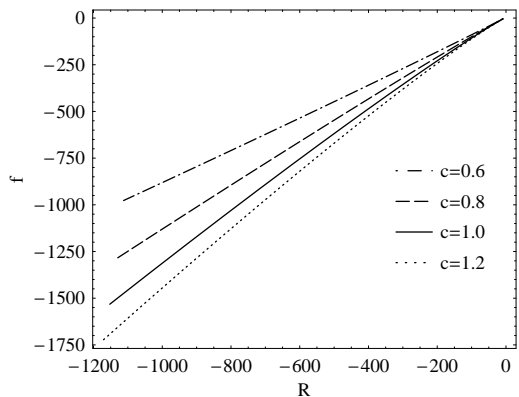} \caption[]{\small
Reconstructed $f(R)$ with $0\leq z\leq10$ and $c=0.6$ (dash-dotted),
$c=0.8$ (dahsed), $c=1.0$ (solid) and $c=1.2$ (dotted).}
\label{fig:fR}
\end{center}
\end{figure}

\begin{figure}[htbp]
\begin{center}
\includegraphics[scale=0.45]{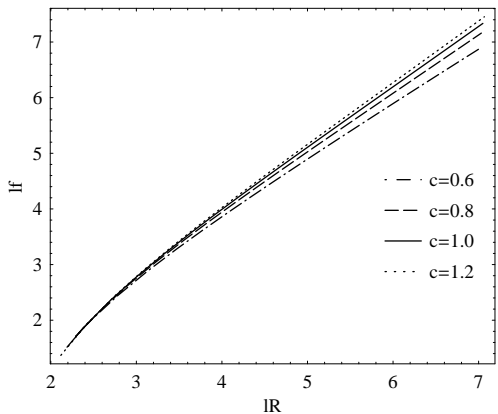} \caption[]{\small
Reconstructed $f(R)$ in $lf-lR$ plane with $0\leq z\leq10$.
$lf=ln(-f)$ and $lR=ln(-R)$. } \label{fig:lfR}
\end{center}
\end{figure}

Once we fix $c$ and $\Omega_{m0}$, the evolution of
$\Omega_\Lambda(z)$ can be determined by solving (\ref{domega}) with
the initial condition (\ref{initial}). Then by (\ref{H(z)}) we can
obtain $H(z)$. Now we assume the holographic dark energy as an
underlying theory of dark energy, and we want to find the
corresponding $f(R)$ theory as an effective description. We follow
the method proposed in\cite{fR recon}. In a FRW universe, the Ricci
scalar can be expressed in terms of Hubble parameter: \eq
R=-6(\dot{H}+2H^2+{k\over a^2}).\label{R} \ee In this paper we set
$k=0$. Note that here we assume the signature as $\{+,-,-,-\}$, the
same as in\cite{fR recon}, and therefore $R$ is always negative (so
it is with $f$). Once we choose $\{-,+,+,+\}$, the minus sign in
front of (\ref{R}) would disappear. This is just a matter of
convention, which means no physical difference. Let's start from the
action \eq S=\int d^4x\sqrt{-g}\left [f(R)+\mathcal{L}_m\right],\ee
where $\mathcal{L}_m$ is the matter Lagrangian. We use the units
$M_p=c=\hbar=1$. Variation with respect to the metric leads to the
modified field equation\cite{derive field eq} \eq G_{\mu \nu} =
R_{\mu \nu} -  \frac{1}{2} R g_{\mu \nu} = T^{(curv)}_{\mu \nu} +
T^{(m)}_{\mu \nu} \label{eq: field eq},\ee where $G_{\mu \nu}$ is
the Einstein tensor, and an effective stress-energy tensor
containing the higher order contributions is defined by
\begin{eqnarray}
T^{(curv)}_{\mu \nu}  &=&  \frac{1}{f'(R)} \left \{ g_{\mu \nu} \left [ f(R) - R f'(R) \right ] /2 \right . \nonumber \\
~ & ~ & \nonumber \\
~ & + & \left . f'(R)^{; \alpha \beta} \left ( g_{\mu \alpha} g_{\nu
\beta} - g_{\mu \nu} g_{\alpha \beta} \right ) \right \} \label{eq:
curvstress}
\end{eqnarray}
and the matter's contribution is in $T^{(m)}_{\mu \nu} =
\tilde{T}^{(m)}_{\mu \nu}/f'(R)$ with $\tilde{T}^{(m)}_{\mu \nu}$
the standard minimally coupled matter stress-energy tensor. With the
FRW metric, we obtain the modified Friedmann equations \eq H^2
+\frac{k}{a^2} = \frac{1}{3} \left [ \rho_{curv} +
\frac{\rho_m}{f'(R)} \right ] \ , \label{eq: fried1}\ee \eq 2
\frac{\ddot{a}}{a} + H^2+\frac{k}{a^2} = - \left ( p_{curv} + p_m
\right ) \ , \label{eq: fried2}\ee and the continuity equation \eq
\dot{\rho}_{tot} + 3 H (\rho_{tot} + p_{tot}) = 0 \ .\ee The three
equations are not independent and we combine them to get one
equation
\begin{eqnarray}
\dot{H} &=& -\frac{1}{2 f'(R)} \left \{ 3 H_0^2 \Omega_M (1 + z)^3 +
\ddot{R} f''(R) \right . \nonumber \\
~ & ~ & \nonumber \\
~ & + & \left . \dot{R} \left [ \dot{R} f'''(R) - H f''(R) \right ]
\right \} \ ,\label{key}
\end{eqnarray}
where the prime denotes derivative with respect to $R$. Using the
relation $d/dt = - (1 + z) H d/dz $ to replace the variable $t$ by
$z$ , (\ref{key}) can be transformed into a third order differential
equation of $f(z)$ \eq {\cal{C}}_3(z) \frac{d^3f}{dz^3} +
{\cal{C}}_2(z) \frac{d^2f}{dz^2} + {\cal{C}}_1(z) \frac{df}{dz}= -3
H_0^2 \Omega_{m0}(1 + z)^3 \label{df3}, \ee where ${\cal{C}}_n(z)$
consists of $H(z)$ and its derivatives. Once we know the function
$H(z)$, the coefficients ${\cal{C}}_n(z)$'s can be calculated.
However, in our case of the holographic dark energy model, $H(z)$
cannot be derived analytically. By (\ref{H(z)}) and (\ref{domega}),
$H(z)$ and its derivatives can be expressed by $\Omega_\Lambda(z)$,
therefore, after painful derivation, the coefficients
${\cal{C}}_n(z)$'s can be ultimately expressed by the combinations
of $\Omega_\Lambda(z)$. Although $\Omega_\Lambda$ itself can be
solved by (\ref{domega}) alone, it has to be considered as a part of
the whole set of  differential equations in order to solve $f(z)$.
So we consider (\ref{domega}) and (\ref{df3}) as a differential
equation set, of which one initial condition is (\ref{initial}).
According to \cite{fR recon}, the other initial conditions are \eq
\left(\frac{df}{dz}\right)_{z=0}=\left(\frac{dR}{dz}\right)_{z=0},\label{df10}\ee
\eq
\left(\frac{d^2f}{dz^2}\right)_{z=0}=\left(\frac{d^2R}{dz^2}\right)_{z=0},\label{df20}\ee
\eq f(z=0)=f(R_0)=6H_0^2(1-\Omega_{m0})+R_0 .\label{f0}\ee Given
$\Omega_{m0}$ and $c$, with these four initial conditions
(\ref{initial})and (\ref{df10})\,-\,(\ref{f0}), the differential
equation set (\ref{domega}) and (\ref{df3}) can be solved
numerically. The reconstructed function $f(R)$ is presented in
Fig.\ref{fig:fR}, where we set $\Omega_{m0}=0.29$ and $c={0.6,\
0.8,\ 1.0,\ 1.2}$ respectively. For the sake of comparison, we also
show the same result on a $lf-lR$ plane in Fig.\ref{fig:lfR}, where
$lf=ln(-f)$ and $lR=ln(-R)$ as used in\cite{fR recon}. Note that we
set the values of $c$ within the range $0.6\leq z\leq 1.2$, which is
consistent with fitting results according to the works in\cite{holo
fit}. Compared with the reconstructed function $f(R)$ for quiessence
and that for Chaplygin gas (Fig.2 and Fig.4 in\cite{fR recon}), we
can see that the three figures are similar. This is because the
values of the parameters are set to be around their best fits. This
is in effect approximately equivalent to reconstruct $f(R)$ with
$H(z)$ directly from observational data. We expect future
observations with more accurate data will discriminate between these
models.

\section{Discussion and conclusion}

The reconstructed $f(R)$ theory is naturally consistent with the
solar system experiment by the reconstruction method. In fact, this
requirement is just the physical motivation for the initial
conditions (\ref{df10}) and (\ref{df20}). For one thing, if we
rewrite (\ref{eq: fried1}) explicitly with $8\pi G$ \eq H^2 =
\frac{8 \pi G}{3} \left [ \rho_{curv} + \frac{\rho_m}{f'(R)} \right
],\ee we can see that $f'(R)$ is effectively modified the Newton
gravitational constant $G$ as $G/f'(R)$, that is , a variable
gravitational coupling. In order to be compatible with solar system
experiments, at $z=0$, we must require $G/f'(R_0)=G$ or $f'(R_0)=1$.
By \eq f'(R_0) = 1 \rightarrow \left [ \left ( \frac{dR}{dz} \right
)^{-1} \frac{df}{dz} \right ]_{z = 0} = 1 .\ee this leads to
(\ref{df10}). For the other, it is shown\cite{solar test} that the
consistency with solar system test also requires $f''(R_0)=0$, which
directly gives rise to (\ref{df20}). As for the remote past, there
is no reason for us to impose this requirement since the experiments
are done today and the validity of the result holds only at $z\sim
0$.

\begin{figure}[htbp]
\begin{center}
\includegraphics[scale=0.45]{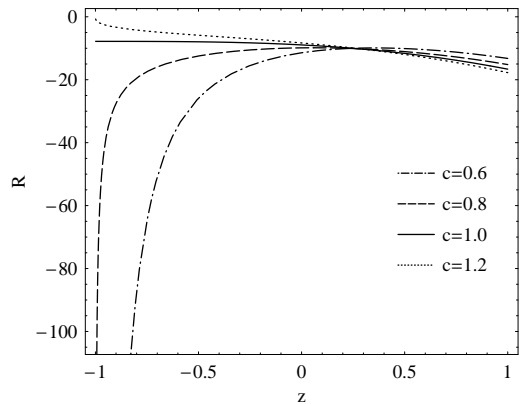}
\caption[]{\small The future evolution of $R$} \label{fig:Rz future}
\end{center}
\end{figure}

\begin{figure}[htbp]
\begin{center}
\includegraphics[scale=0.45]{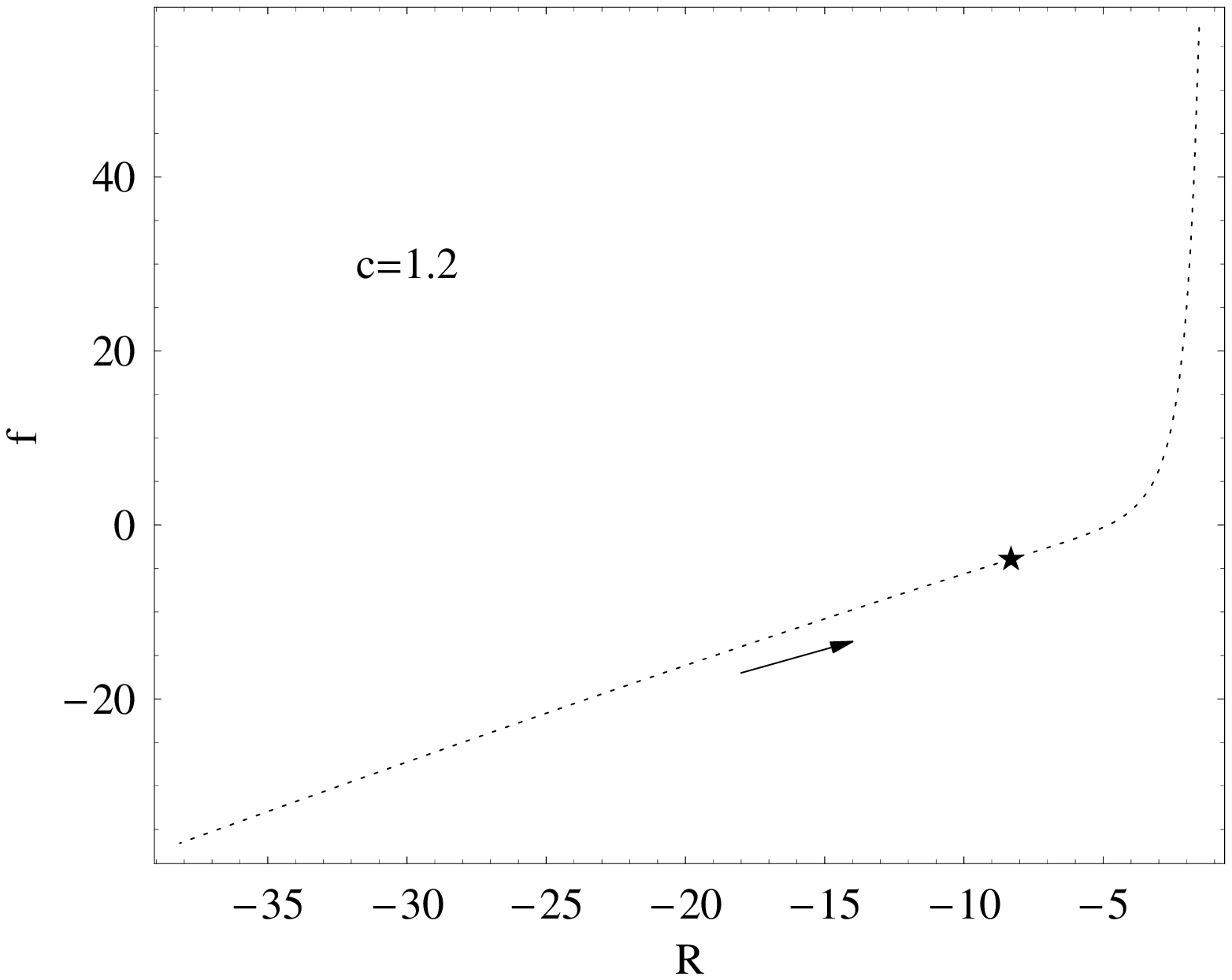}
\includegraphics[scale=0.45]{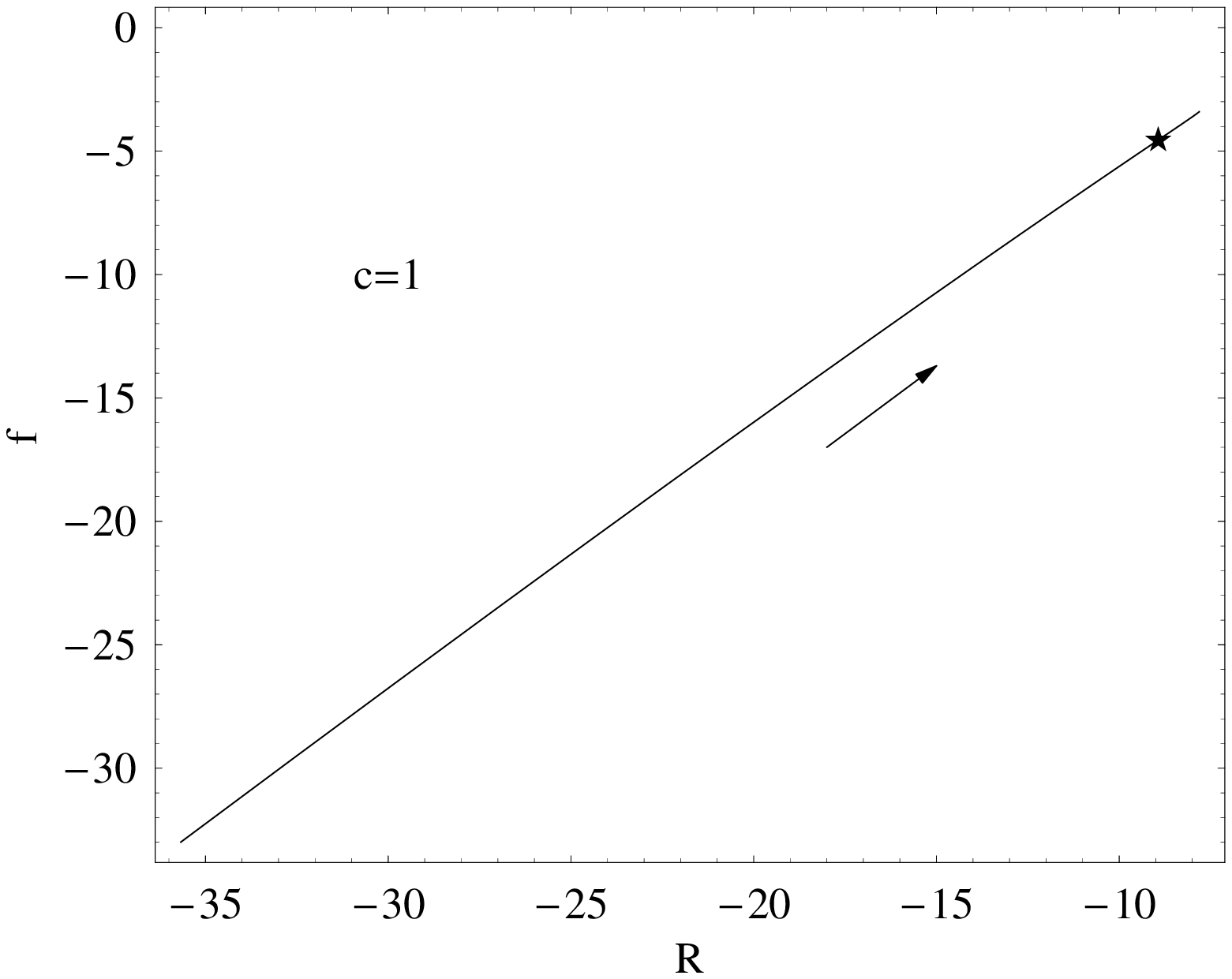}
\includegraphics[scale=0.45]{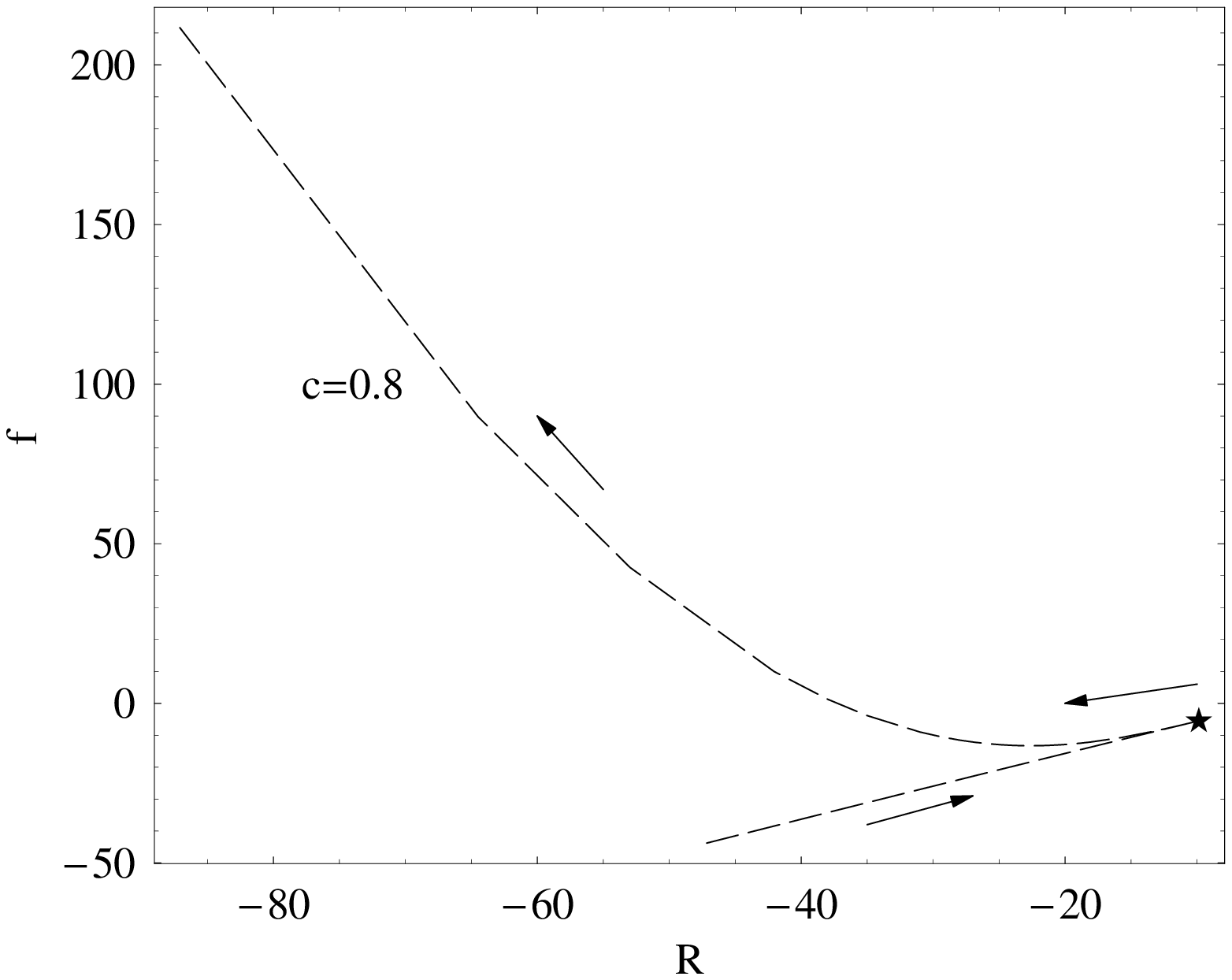}
\caption[]{\small Reconstructed $f(R)$. The curves are plotted with
$z$ from around $2$ down to $-1$. The arrow denotes the direction of
$z$ increasing. The star denotes current value at $z=0$. The
turnaround point is the point where the total energy density starts
to increase after phantom energy dominates over matter.}
\label{fig:fR future}
\end{center}
\end{figure}

Fig.\ref{fig:fR} shows that for small $|R|$ (small $z$ also), the
functions $f(R)$ are indistinguishable for different parameter $c$.
As we mentioned before, $c$ is a crucial parameter characterizing
the nature of the holographic dark energy. Differences between the
corresponding $f(R)$ functions become significant as $|R|$ (or $z$)
increases. To further illustrate that the reconstructed theory does
reflect the distinctive effect of $c$, we consider the future
evolution scenario. Fig.\ref{fig:Rz future} shows the future
evolution of $R$. As is expected, for $c<1$, the curves indicate the
typical phantom behavior: $|R|\rightarrow\infty$. This is because
the dark energy with EoS$<-1$ dominates over matter and the phantom
energy density increases with time, tears apart structures and a Big
Rip is unavoidable. For $c=1$, $|R|$ varies little and the dark
energy becomes more and more like a cosmological constant. For
$c>1$, $R$ vanishes in the future. In Fig.\ref{fig:fR future}, we
can see that the difference is more distinctively reflected by the
function $f(R)$ reconstructed based upon the future evolution of the
holographic dark energy model. For $c=1.2$, as $R$ approaches zero,
$f$ increases from negative to positive, which may indicate a
inverse power law dependence of $f$ on $R$. This is consistent with
the models proposed in \cite{Carroll} and \cite{R-nRm}.  For $c=1$,
the straight line manifests a linear dependence on $R$ up to a
constant, which is consistent with the de Sitter phase where
$f=R+2\Lambda$. For $c=0.8$, the curve first meets a turnaround
point, at which the decreasing $|R|$ begins to increase due to the
domination of the phantom-like dark energy. It can be checked that
in this case, the turnaround redshift is in the near future for
$c=0.8$ while it is in the near past for $c=0.6$, namely, the
domination of phantom-like dark energy begins earlier for smaller
$c$. As the universe evolves, $|R|$ keeps growing, and $f$ decrease
first and then increases to become positive. Both $f$ and $R$ become
divergent in the final Big Rip. Further analysis shows that for
$c=0.6$, $f\rightarrow-\infty$; for $c=0.8$, $f\rightarrow+\infty$,
which clearly reveals that the form of the function $f(R)$ may be
significantly different for different $c$. We note that the
existence of the turnaround point is a universal feature for all the
phantom-dark energy models realized in $f(R)$ theories, due to the
competition between dark energy and matter.

In conclusion, we have reconstructed the function $f(R)$ in the
extended theory of gravity according to the holographic dark energy.
The basic reconstruction procedure can be simply summarized as:
first, express $R$ as a function of $z$; then $f$ can also be
considered as a function of $z$ by $f(z)=f[R(z)]$; thirdly obtain a
third order differential equation for $f(z)$ and solve it with some
initial conditions; and finally reconstruct $f(R)$ from $R(z)$ and
$f(z)$. Note that in this procedure, the Hubble parameter $H(z)$ and
its derivatives with respect to $z$ enter into $R$ and the
coefficients of the differential equation for $f(z)$. Since $H(z)$
and its derivatives can be expressed by $\Omega_\Lambda$, what we
are dealing with can be treated essentially as a differential
equation set with the unknown functions $f(z)$ and
$\Omega_\Lambda(z)$ to be solved. With some initial conditions
imposed by physical consideration, we solved the differential
equation set and found the function $f(R)$ numerically.

In addition, it should be emphasized that the holographic dark
energy is the result obtained within the framework of general
relativity, rather than any other extended theory such as $f(R)$
theory. What we have done is to reconstruct the $f(R)$ theory which
effectively describes the holographic dark energy in Einstein
gravity. Whether the holographic vacuum energy can be generalized to
$f(R)$ theories and what it looks like are questions worth further
investigation.

\section*{Acknowledgments}
We thank Rong-Gen Cai for helpful discussion. This work was
supported by NSFC under Grant No. 10533010, and SRF for ROCS, SEM of
China.

\appendix*

\section{}
Here we list the some expressions essential for practical
calculation. The coefficients in (\ref{df3}) are
\begin{eqnarray}
{\cal{C}}_1 & = & \dot{R}^2 \left ( \frac{dR}{dz} \right )^{-4}
\left [ 3 \left ( \frac{dR}{dz} \right )^{-1}
\left ( \frac{d^2R}{dz^2} \right )^{2} - \frac{d^3R}{dz^3} \right ] \nonumber \\
~ & ~ & - \left ( \ddot{R} - \dot{R} H \right ) \left ( \frac{dR}{dz} \right )^{-3} \frac{d^2R}{dz^2} \nonumber \\
~ & ~ & -2 (1 + z) H \frac{dH}{dz} \left ( \frac{dR}{dz} \right
)^{-1} , \label{eq: h1}
\end{eqnarray}
\begin{equation}
{\cal{C}}_2 = \left ( \ddot{R} - \dot{R} H \right ) \left (
\frac{dR}{dz} \right )^{-2} - 3 \dot{R}^2 \left ( \frac{dR}{dz}
\right )^{-4} \frac{d^2R}{dz^2}  , \label{eq: h2}
\end{equation}
\begin{equation}
{\cal{C}}_3 = \dot{R}^2 \left ( \frac{dR}{dz} \right )^{-3}  .
\label{eq: h3}
\end{equation}
$R(z)$ and its derivatives are
\begin{equation}
R = -6 \left [ 2 H^2 - (1 + z) H \frac{dH}{dz} \right ]  .
\label{eq: rvsh}
\end{equation}
\begin{equation}
\frac{dR}{dz} = -6 \left \{ -(1 + z) \left ( \frac{dH}{dz} \right
)^2 + H \left [ 3 \frac{dH}{dz} - (1 + z) \frac{d^2H}{dz^2} \right ]
\right \} \label{eq: drdz}
\end{equation}
\begin{equation}
\dot{R} = - (1 + z) H \frac{dR}{dz}  , \label{eq: dotR}
\end{equation}
\begin{eqnarray}
\ddot{R} - \dot{R} H & = & 6 (1 + z) H^2 \left \{ 3 (1 + z)^2 \frac{dH}{dz} \frac{d^2H}{dz^2} +\right . \nonumber \\
~ & ~ & + \left . H \left [ (1 + z)^2 \frac{d^3H}{dz^3}  - 6
\frac{dH}{dz} \right ] \right \}  . \label{eq: ddotR}
\end{eqnarray}
Higher order derivatives of $H(z)$ are too complicated to be listed
here. In practical calculation, we use computer program for
derivation. In addition, the code for $Mathematica\ 5.0$ we used is
available on request.


\begin{thebibliography}{99}

\bibitem{SN}
A.~G.~Riess {\it et al.},
Astron.\ J.\  {\bf 116} (1998) 1009 [astro-ph/9805201];\\
S.~Perlmutter {\it et al.}, Astrophys.\ J.\  {\bf 517} (1999) 565
[astro-ph/9812133].

\bibitem{CMB} 

D.~N.~Spergel {\it et al.}, [astro-ph/0603449];\\
L.~Page {\it et al.}, [astro-ph/0603450];\\
G.~Hinshaw {\it et al.}, [astro-ph/0603451];\\
N.~Jarosik {\it et al.}, [astro-ph/0603452].

\bibitem{LSS}
M.~Tegmark {\it et al.},  Phys.\ Rev.\ D {\bf 69} (2004) 103501  [astro-ph/0310723];\\
M.~Tegmark {\it et al.},  Astrophys.\ J.\  {\bf 606} (2004) 702  [astro-ph/0310725];\\
U.~Seljak {\it et al.}, Phys.\ Rev.\ D {\bf 71} (2005) 103515  [astro-ph/0407372];\\
J.~K.~Adelman-McCarthy {\it et al.}  [SDSS Collaboration],  Astrophys.\ J.\ Suppl.\  {\bf 162} (2006) 38  [astro-ph/0507711];\\
K.~Abazajian {\it et al.}, [astro-ph/0410239]; [astro-ph/0403325]; [astro-ph/0305492];\\
M.~Tegmark {\it et al.}, Phys.\ Rev.\  D {\bf74} (2006) 123507
[astro-ph/0608632].

\bibitem{0603057}
E.~J.~Copeland, M.~Sami and S.~Tsujikawa,
  Int.\ J.\ Mod.\ Phys.\  D {\bf 15} (2006) 1753
  [hep-th/0603057].

\bibitem{cc}
P.~J.~E.~Peebles and B.~Ratra,
 Rev.\ Mod.\ Phys.\  {\bf 75} (2003) 559  [astro-ph/0207347];\\
S.~M.~Carroll,
  Living Rev.\ Rel.\  {\bf 4} (2001) 1
  [astro-ph/0004075];\\
S.~Weinberg, Rev.\ Mod.\ Phys.\  {\bf 61} (1989) 1.

\bibitem{holo}
G.~'t Hooft, [gr-qc/9310026];\\
L.~ Susskind,  J. Math. Phys. 36 (1995) 6377 [hep-th/9409089]

\bibitem{Cohen}
A.~Cohen, D.Kaplan and A.~Nelson, Phys.\ Rev.\ Lett. {\bf 82} (1999)
4971; [hep-th/9803132]

\bibitem{Hsu}
S.~D.~H.~Hsu, hep-th/0403052

\bibitem{Li}
M.~Li, Phys.\ Lett.\  B {\bf 603} (2004) 1  [hep-th/0403127].

\bibitem{holo fit}

Q.~G.~Huang and Y.~G.~Gong,
   JCAP {\bf 0408} (2004) 006
  [astro-ph/0403590];\\
X.~Zhang and F.~Q.~Wu,
   Phys.\ Rev.\  D {\bf 72} (2005) 043524
  [astro-ph/0506310];\\
Z.~Chang, F.~Q.~Wu and X.~Zhang,
  Phys.\ Lett.\  B {\bf 633} (2006) 14
  [astro-ph/0509531];\\
Z.~L.~Yi and T.~J.~Zhang,
  Mod.\ Phys.\ Lett.\  A {\bf 22} (2007) 41
  [astro-ph/0605596];\\
X.~Zhang and F-Q.~Wu, Phys.\ Rev.\ D {\bf 76} (2007) 023502
[astro-ph/0701405].

\bibitem{effective action QG}
I.L. Buchbinder, S.D. Odintsov, I.L. Shapiro, {\it Effective Action
in Quantum Gravity}, IOP Publishing (1992) Bristol;\\
N.D. Birrell, P.C.W. Davies, {\it Quantum Fields in Curved Space},
Cambridge Univ. Press, Cambridge (1982);\\
R.~Utiyama and B.~S.~De Witt J. Math. Phys. {\bf 3} (1962) 608.

\bibitem{effective action string}
M.~Green, J.~Schwarz and E.~Witten, Superstring Theory, Cambridge
Univ. Press, Cambridge (1987);\\
A.~A.~Tseytlin and C.~Vafa, Nucl. Phys. B {\bf 372} (1992) 443;\\
G.~Veneziano, Phys. Lett. B {\bf 265} (1991) 287;\\
M.~Gasperini, J.~Maharana and G.~Veneziano, Phys. Lett. B {\bf 272}
(1991) 277;\\
K.~A.~Meissner and G.~Veneziano, Phys. Lett. B {\bf 267} (1991) 33.




\bibitem{Starobinsky}
A.~A.~Starobinsky, Phys. lett. B {\bf 91} (1980) 99.


\bibitem{Carroll}
S.~M.~Carroll, V.~Duvvuri, M.~Trodden and M.~S.~Turner, Phys. Rev. D
{\bf 70} (2003) 043528

\bibitem{R-nRm}
S.~Nojiri and S.~D.~Odintsov, Phys. Rev. D {\bf 68} (2003) 123512


\bibitem{conformal}
P.~Teyssandier  J. Math. Phys. {\bf 24}(1983) 2793;\\
K.~Maeda, Phys. Rev.  D, {\bf 39} (1989) 3159;\\
D.~Wands, Class. Quant. Grav.  {\bf 11} (1994) 269;\\
S.~Capozziello, R.~de Ritis and A.~A.~Marino. Gen. Rel. Grav. {\bf
30} (1998) 1247;\\
S.~Capozziello and G.~Lambiase, Gen. Rel. Grav. {\bf 32} (2000) 295;\\
S.~Gottl\"ober, H.-J.~Schmidt and A.~A.~Starobinsky, Class. Quant.
Grav. {\bf 7} (1990) 893.


 \bibitem{fR recon}
S.~Capozziello, V.~F.~Cardone and A.~Troisi, Phy. Rev. D {\bf 71}
(2005) 0403503

\bibitem{holo quintom}
X.~Zhang, Phys. Rev. D {\bf 74} (2006) 103505 [astro-ph/0609699].

\bibitem{holo quintessence}
X.~Zhang, Phys. Lett. B {\bf 648} (2007) 1 [astro-ph/0604484].

\bibitem{holo tachyon}
J.~ Zhang, X.~Zhang and H.~Liu, Phys. Lett. B {\bf 651} (2007) 84
arXiv:0706.1185[astro-ph].

\bibitem{holo hessence}
W.~Zhao arXiv:0706.2211[astro-ph].

\bibitem{Palatini}
M. Ferraris, M. Francaviglia, I. Volovich, Class. Quant. Grav. {\bf
11} (1994) 1505; \\
G. Allemandi, A. Borowiec, M. Francaviglia, Phys. Rev. D, {\bf 70}
(2004) 043524; \\
G. Allemandi, A. Borowiec, M. Francaviglia, Phys. Rev. D, {\bf 70}
(2004) 103503.


\bibitem{ACCF}
G. Allemandi, M. Capone, S. Capozziello, M. Francaviglia,
[hep-th/0409198].


\bibitem{derive field eq}
S.~Capozziello, Int. J. Mod. Phys. D {\bf 11} (2002) 483;\\
S.~Capozziello, S.~Carloni and A.~Troisi, [astro-ph/0303041].


\bibitem{solar test}
R. Dick, Gen. Rel. Grav., {\bf 36} (2004) 217; \\
A.E. Dominguez, D.E. Barraco, Phys. Rev. D, {\bf 70} (2004) 043505.

\end{thebibliography}
\end{document}